\documentclass[twocolumn,showpacs,preprintnumbers,superscriptaddress,amsmath,amssymb]{revtex4}
\usepackage{graphicx}
\usepackage{dcolumn}
\usepackage{bm}
\usepackage{color}
\usepackage{booktabs, longtable}
\usepackage{multirow}
\usepackage{CJK}
\usepackage{ltxtable}
\usepackage{ulem}
\begin{document}

\title{BCS-BEC crossover in nuclear matter with the relativistic Hartree-Bogoliubov theory}

\author{Ting Ting Sun}
 \affiliation{School of Physics and State Key Laboratory of Nuclear Physics and Technology, Peking University, 100871 Beijing, People's Republic of China}
\author{Bao Yuan Sun}
 \affiliation{School of Nuclear Science and Technology, Lanzhou University, 730000 Lanzhou, People's Republic of China}
 \affiliation{Research Center for Nuclear Physics (RCNP), Osaka University, Ibaraki, 567-0047 Osaka, Japan}
 \affiliation{School of Physics and State Key Laboratory of Nuclear Physics and Technology, Peking University, 100871 Beijing, People's Republic of China}
\author{Jie Meng}
\email{mengj@pku.edu.cn}
 \affiliation{School of Physics and State Key Laboratory of Nuclear Physics and Technology, Peking University, 100871 Beijing, People's Republic of China}
 \affiliation{School of Physics and Nuclear Energy Engineering, Beihang University, 100191 Beijing, People's Republic of China}
 \affiliation{Department of Physics, University of Stellenbosch, 7602 Stellenbosch, South Africa}

\date{\today}

\begin{abstract}
Based on the relativistic Hartree-Bogoliubov theory, the influence of the pairing interaction strength on the di-neutron correlations and the crossover from superfluidity of neutron Cooper pairs in the $^{1}S_{0}$ channel to Bose-Einstein condensation of di-neutron pairs is systematically investigated in the nuclear matter. The bare nucleon-nucleon interaction Bonn-B is taken in the particle-particle channel with an effective factor to simulate the medium effects and take into account the possible ambiguity of pairing force, and the effective interaction PK1 is used in the particle-hole channel. If the effective factor is larger than $1.10$, a di-neutron BEC state appears in the low-density limit, and if it is smaller than $0.85$, the neutron Cooper pairs are found totally in the weak coupling BCS region.
The reference values of several characteristic quantities which characterize the BCS-BEC crossover are obtained respectively from the dimensionless parameter $1/(k_{\rm Fn}a)$ with $a$ the scattering length and $k_{\rm{Fn}}$ the neutron Fermi momentum, the zero-momentum transfer density correlation function $D(0)$ and the effective chemical potential $\nu_{\rm n}$.
\end{abstract}

\pacs{21.65.-f, 21.30.Fe, 21.60.Jz, 74.20.Fg, 03.75.Nt}
\maketitle

\section{Introduction}\label{Chapter1}

The BCS-BEC crossover, which is described as the evolution of the pairing phenomenon from the weakly coupled Bardeen-Cooper-Schrieffer (BCS) type to the strongly correlated Bose-Einstein condensation (BEC) state with increasing pairing interaction,
is one of the important issues in the study of the pairing correlations in Fermion systems.
The transition from BCS to BEC was investigated theoretically in several domains of physics, such as in excitonic semiconductors~\cite{EaglesPhysRev1969}, superconductors~\cite{LeggettJPhysC1980} and attractive Fermi gases~\cite{NozieresJLTP1985}. Although BCS and BEC are physically two quite different limits, it was found that the evolution between them is smooth and continuous~\cite{LeggettJPhysC1980, NozieresJLTP1985, OhashiPRL2002}.

Since the strength of inter-atomic interaction can be controlled via a magnetically driven Feshbach resonance, ultracold Fermi atomic gases provide a special laboratory for studying strongly correlated Fermion systems. Recently, the BCS-BEC crossover phenomenon has been experimentally realized in ultracold quantum atomic gas~\cite{GreinerNature2003, JochimScience2003, ZwierleinPRL2003}.

In nuclear physics, however, the BCS-BEC crossover phenomenon is still a challenging topic because of the impossibility to control the nuclear pairing force directly. Indirectly, one can change the strength of pairing interaction in nuclear matter by altering density or temperature. As a result, density and temperature are treated as important parameters in the study of nuclear pairing correlations.

For the neutron-proton pairing, the BCS-BEC crossover has been investigated in the $^{3}S_{1}-^{3}D_{1}$ channel, in which the strong spatial correlation and the BEC of the deuterons may occur at low densities~\cite{AlmNPA1993, SteinZPhysA1995,BaldoPRC1995,LombardoPRC2001}.

For the neutron-neutron pairing, the correlation is expected to be significant in low-density nuclear matter as well. It is well known that the bare neutron-neutron interaction in the $^1S_0$ channel leads to a virtual state around zero energy characterized by a large negative scattering length $a\approx-18.5\pm0.4~{\rm fm}$~\cite{TeramondPRC1987}, implying a very strong attraction between two neutrons in the spin singlet state. Furthermore, theoretical predictions suggest that around $1/10$ of the normal nuclear density $\rho_{0}$, the $^1S_0$ pairing gap may take a considerably larger value than that around $\rho_0$~\cite{BaldoNPA1990,TakatsukaPTPS1993,DeanRevModPhys2003}. In addition, the strong di-neutron correlations are also supported by the enhancement of two-neutron transfer cross sections in several heavy nuclei~\cite{WvonOertzenRPP2001}. In the weakly bound neutron-rich nuclei, di-neutron correlations are enhanced due to the couplings with the continuum and play an important role for unstable nucleus and the formation of nuclear halo~\cite{MengLiPRL1996,MengRCHBNPA1998,MengPInteractionPRC1998}. Recently, di-neutron emission in $^{16}$Be was observed for the first time, indicating the structure of di-neutron clusters inside neutron-rich nuclei~\cite{SpyrouPRL2012}.

The progress in both theoretical and experimental investigations on di-neutron correlations in weakly bound nuclei has stimulated lots of interest in searching for possible BCS-BEC crossover of neutron pairing. The influences on di-neutron correlations by the Skyrme-type density dependent force and the finite-range Gogny force are discussed in Ref.~\cite{MengPInteractionPRC1998}. In nuclear matter, the study of the BCS-BEC crossover phenomenon was mainly performed by using the finite-range Gogny interaction and the zero-range contact interaction~\cite{MatsuoPRC2006, MargueronSagawaPRC2007, IsayevPRC2008, MaoPRC2009}. It has been found that the di-neutron correlations get stronger as density decreases, and BCS-BEC crossover could occur at low densities. Besides, based on the bare force given by a superposition of three Gaussian functions~\cite{MatsuoPRC2006} or the bare nucleon-nucleon interaction Bonn potential~\cite{SunPLB2010}, the BCS-BEC crossover in the nuclear matter has also been investigated. It was shown that the spatial structure of the wave function for the neutron Cooper pairs evolves from BCS-type to BEC-type as density decreases. From several characteristic quantities, such as the effective chemical potential, the quasi-particle excitation spectrum and the density correlation function, there is no evidence for the BEC state of di-neutron pairs at any density. However, it was argued that a Bose  di-neutron  gas may be found in the low-density limit in quasi-two-dimensional neutron systems as long as the slab is thin enough~\cite{YoshikoQ2DPRC2009}. Recently, the study on the BCS-BEC crossover phenomenon has been extended to finite temperature~\cite{HuangPRC2010}. The thermodynamic signal from the temperature dependence of specific heat suggests that BCS-BEC crossover has been found in the low-density region. In finite nuclei, the coexistence of BCS- and BEC-like spatial structures of di-neutron wave functions has also been revealed in the halo nucleus $^{11}$Li~\cite{HaginoPRL2007}. From the di-neutron wave function, a strong correlation between the valence neutrons appears on the surface of the nucleus.

For most investigations on di-neutron correlations, for convenience, the phenomenological effective nuclear forces are used in the particle-particle ($pp$) channel. The effective interactions in the relativistic mean-field (RMF) theory are also used in the $pp$ channel~\cite{KucharekZPhysA1991,LiJIJMPE2008}. However, one has to introduce an effective factor for this kind of pairing interaction in order to obtain reasonable values for the gap parameter~\cite{LiJIJMPE2008}.
In fact, details of the effective nucleon-nucleon interaction in nuclear matter and finite nuclei are as yet not completely clarified, since most of the experimental data are not very sensitive to their details. It is generally believed that the bare nucleon-nucleon interaction should be corrected by the medium polarization effects ( also referred to as the screening effects) in extracting the effective interactions in $pp$ channel at various densities~\cite{CaoPRC2006}. Therefore, it is of great interest to study the behavior of di-neutron correlations and the BCS-BEC crossover phenomenon with different strengths of the pairing interaction.

As the RMF theory has achieved lots of success in the descriptions of both nuclear matter and finite nuclei near or far from the stability line~\cite{SerotANP1986,RingPPNP1996, MengPPNP2006}, following a previous investigation~\cite{SunPLB2010}, the influence of the strength of pairing interaction on di-neutron correlations in the $^{1}S_{0}$ channel will be studied with the relativistic Hartree-Bogoliubov (RHB) theory in nuclear matter in this work. The bare nucleon-nucleon interaction, i.e., the relativistic Bonn potential~\cite{MachleidtANP1989}, will be adopted in the $pp$ channel with an effective factor to simulate the medium polarization effects and take into account the possible ambiguity of pairing force~\cite{CaoPRC2006}, and the relativistic mean-field model is used in the particle-hole ($ph$) channel.
The pairing strength marking the crossover from the strong correlated BEC state to the weakly correlated BCS state will be discussed. Section ~\ref{Chapter2} briefly introduces the formulism of the RHB theory for nuclear matter. The results are given and discussed in Section~\ref{Chapter3} and finally the work is summarized in Section~\ref{Chapter4}.

\section{Theoretical Framework}\label{Chapter2}

The RMF model with nonlinear $\sigma$- and $\omega$-meson self-coupling is employed in the $ph$ channel to describe the bulk properties of the nuclear matter~\cite{SerotANP1986,RingPPNP1996, MengPPNP2006}. For the static and uniform infinite nuclear matter, the Coulomb field is neglected and
the space-like components as well as the differential of the time-like components of the meson fields vanish.

In the RHB theory, the meson fields are treated dynamically beyond the mean-field approximation to provide the pairing field via the anomalous Green's functions, which gives a uniform description for the mean field and the pairing field~\cite{KucharekZPhysA1991}. In the case of infinite nuclear matter, the RHB equation is reduced to the usual BCS equation. The BCS ground state is defined as
 \begin{equation}
 |{\rm BCS}\rangle=
 \prod_{k>0}(u_{k}+v_{k}\hat{a}_{k\uparrow}^{\dag}\hat{a}_{-k\downarrow}^{\dag})|-\rangle,
  \label{Eq:BCSG.S.}
 \end{equation}
where $u_{k}$ and $v_{k}$ represent the BCS variational parameters, and $\hat{a}_{k \uparrow}^{\dag}$ ($\hat{a}_{-k \downarrow}^{\dag}$) are creation operators of a particle with momentum $\bm{k}$ ($-\bm{k}$) and spin $\uparrow$ ($\downarrow$) on top of the vacuum $|-\rangle$~\cite{BardeenBCS2PhysRev1957}.

For the $^1S_0$ channel, the pairing gap function $\Delta(p)$ satisfies
\begin{equation}
    \Delta(p) = -\frac{1}{(2\pi)^{3}}\int \lambda\cdot v_{pp}(\bm{k},{\bm p})\frac{\Delta(k)}{2E_k}d{\bm k},
    \label{Eq:GapEq}
\end{equation}
where $v_{pp}({\bm k}, {\bm p})$ is the matrix element of nucleon-nucleon interaction in the momentum space. Here an effective pairing interaction factor $\lambda$ is introduced to control the pairing interaction strength. The quasi-particle energy $E_k$ can be written as,
 \begin{equation}
 E_k = \sqrt{(\varepsilon_k - \mu)^2 + \Delta(k)^2},
 \label{Eq:q.p.energy}
 \end{equation}
with the single-particle energy $\varepsilon_k$ and the chemical potential $\mu$. The corresponding normal and anomalous density
distribution functions are given in the following forms,
 \begin{equation}
   \rho_k = \frac{1}{2} \left[ 1 - \frac{\varepsilon_k - \mu}{E_k}\right],
   \quad\kappa_k = \frac{\Delta(k)}{2E_k}.
   \label{Eq:rho.kappa}
 \end{equation}

The single-particle energy $\varepsilon_k$ follows from the RMF theory,
 \begin{equation}
 \varepsilon_k = \Sigma_0 + \sqrt{k^2 + M^{\ast2}},
 \label{Eq:s.p.energy}
 \end{equation}
where $\Sigma_{0}$ is the vector potential, $\Sigma_{S}$ the scalar potential, M the nucleon mass, and $M^\ast=M+\Sigma_S$ the effective mass,
\begin{equation}
   \Sigma_S = g_\sigma\sigma,\quad
   \Sigma_0 = g_\omega\omega_0+g_\rho\tau_3\rho_{0,3},
   \label{Eq:vec.sca.potential}
 \end{equation}
with $g_{\sigma}$, $g_{\omega}$ and $g_{\rho}$ the meson-nucleon coupling constants of the respective meson fields $\sigma$, $\omega_{0}$ and $\rho_{0,3}$, and $\tau_{3}$ the third component of the nucleon isospin.

With the mean-field approximation, the meson fields are replaced by
their mean values, and could be solved from the corresponding
equations of motion with the given nucleon densities,
 \begin{eqnarray}
    m^2_\sigma\sigma &=& -g_\sigma \rho_s-g_2\sigma^2-g_3\sigma^3,\\
    m^2_\omega\omega_0 &=& g_\omega \rho_b-c_3\omega_0^3,\\
    m^2_\rho\rho_{0,3} &=& g_\rho \rho_{b,3},
    \label{Eq:meson.eq}
  \end{eqnarray}
where $\rho_s$, $\rho_b$ and $\rho_{b,3}$ are respectively the
scalar density, vector density and isospin vector density,
 \begin{eqnarray}
   \rho_s &=& \frac{1}{\pi^2}\sum_{i=n,p}\int_0^{\infty}\frac{M^\ast_{i}}{\sqrt{k^2+M^{\ast2}_{i}}}\rho_{k,i}k^2dk,\label{Eq:rhos}\\
   \rho_b &=& \rho_n+\rho_p = \frac{1}{\pi^2}\sum_{i=n,p}\int_0^{\infty}\rho_{k,i}k^2dk,\label{Eq:rhob}\\
   \rho_{b,3} &=& \frac{1}{\pi^2}\sum_{i=n,p}\int_0^{\infty}\tau_3\rho_{k,i}k^2dk.\label{Eq:rhob3}
 \end{eqnarray}
The nucleon Fermi momentum $k_{{\rm F}i}~(i=n, p)$ is defined by the nucleon densities $\rho_{i}$ with $\rho_i\equiv k_{{\rm F}i}^3/3\pi^2$. Therefore, for nuclear matter with given baryonic density $\rho_b$ and isospin asymmetry $\zeta=(\rho_n-\rho_p)/\rho_b$, the above equations can be solved by a self-consistent iteration method with no-sea approximation. Accordingly, the chemical potential $\mu$ is obtained via solving Eq.~(\ref{Eq:GapEq}), Eq.~(\ref{Eq:rho.kappa}) and Eq.~(\ref{Eq:rhob}) by the self-consistent iteration.

The relativistic Bonn potential is used in the $pp$ channel, which has a proper momentum behavior determined by the scattering data up to high energies~\cite{MachleidtANP1989}. It is defined as the sum of one-boson-exchange (OBE) potential of the six bosons $\phi=\sigma, \omega, \pi, \rho, \eta, \delta$. The vector meson $\rho$ includes the vector coupling channel $\rho^{V}$, the tensor coupling channel $\rho^{T}$ and the vector-tensor coupling channel $\rho^{VT}$, and the pseudo-scalar mesons $\pi$ and $\eta$ include the pseudo-scalar coupling channel $\pi^{PV}$ and $\eta^{PV}$, respectively.

The matrix element $v_{pp}(\bm{k},\bm{p})$ is
 \begin{equation}
  v_{pp}(\bm{k},\bm{p}) =
  \sum_\phi\frac{\eta_\phi}{2\varepsilon^\ast_k\varepsilon^\ast_p}A_\phi(\bm{k},\bm{p})D_\phi(\bm{q}^2)F_\phi^2(\bm{q}^2),
 \label{Eq:Vpp}
 \end{equation}
where $\varepsilon^{*}_{k}$ is the effective single-particle energy
 \begin{equation}
  \varepsilon_{k}^{*}=\sqrt{k^{2}+M^{*2}}.
  \label{Eq:eff.s.p.energy}
 \end{equation}
$D_\phi(\bm{q}^2)$ is a meson propagator with the momentum transfer $\bm{q}=\bm{k}-\bm{p}$, and $F_\phi(\bm{q}^2)$ is a form
factor in order to get the reasonable value for the
pairing gap,
 \begin{equation}
   D_\phi(\bm{q}^2) = \frac{1}{\bm{q}^2+m_\phi^2},\quad
   F_\phi(\bm{q}^2) = \frac{\Lambda_\phi^2 - m_\phi^2}{\bm{q}^2 + \Lambda_\phi^2},
 \end{equation}
with the meson mass $m_\phi$ and the cutoff parameter $\Lambda_\phi$. The vertex functions of the OBE potential $\eta_{\phi}$ and $A_{\phi}(\bm{k},\bm{p})$ are listed in Table~\ref{Tab:BonnB}. For the $^1S_0$ pairing channel, the matrix element $v_{pp}(k,p)$  is related to $v_{pp}(\bm{k}, \bm{p})$ by the integral,
 \begin{equation}
 v_{pp}(k,p) = \int_0^\pi v_{pp}(\bm{k}, \bm{p})\sin\theta d\theta,
 \end{equation}
 where $\theta$ is the angle between the vectors ${\bm {k}}$ and ${\bm {p}}$.

 \begin{table*}
 \caption{The vertex functions $\eta_{\phi}$ and $A_{\phi}(\bm{k},\bm{p})$ of the OBE potential in the relativistic Bonn-B potential with the corresponding parameters $g_{\sigma}^{B}, g_{\delta}^{B}, g_{\omega}^{B}, g_{\rho}^{B}, f_{\rho}^{B}, f_{\pi}^{B}$ and $f_{\eta}^{B}$. The general expression could be found in Ref.~\cite{MachleidtCNP}. See the text for details. }
 \label{Tab:BonnB}
 \centering
 \begin{tabular}{c|ccccc}
 \hline\hline
 mesons        &~ $\phi$      &~~& $\eta_\phi$                             &~~  & $A_\phi(\bm{k},\bm{p})$\\ \hline
               &~ $\sigma$    &~~& $-(g_\sigma^{B})^{2}$                           &~~  &$M^{\ast2}+\varepsilon^\ast_k\varepsilon^\ast_p-\bm{k}\cdot\bm{p}$\\
 \raisebox{2.3ex}[0pt]{scalar}
               &~ $\delta$    &~~& $-(g_\delta^{B})^{2}$                           &~~  &$M^{\ast2}+\varepsilon^\ast_k\varepsilon^\ast_p-\bm{k}\cdot\bm{p}$\\ \hline
               &~ $\omega$    &~~& $(g_\omega^{B})^{2}$                           &~~  &$2\left( 2\varepsilon^\ast_k\varepsilon^\ast_p-M^{\ast2} \right)$\\
               &~ $\rho^V$    &~~& $(g_\rho^{B})^{2}$                             &~~  &$2\left( 2\varepsilon^\ast_k\varepsilon^\ast_p-M^{\ast2} \right)$\\
 \raisebox{2.3ex}[0pt]{vector}
               &~ $\rho^T$    &~~& $\left( f_\rho^{B}/2M \right)^2$           &~~  &$\left( M^{\ast2}+3\varepsilon^\ast_k\varepsilon^\ast_p+\bm{k}\cdot\bm{p}\right)\bm{q}^2$\\
               &~ $\rho^{VT}$ &~~&$\left( f_\rho^{B} g_\rho^{B}/M \right) M^\ast$ &~~  &$\bm{q}^2$\\ \hline
               &~ $\pi^{PV}$  &~~&$\left( f_\pi^{B}/m_\pi \right)^2$          &~~  &$\left( M^{\ast2}+\varepsilon^\ast_k\varepsilon^\ast_p+\bm{k}\cdot\bm{p}\right)\bm{q}^2$\\
 \raisebox{2.3ex}[0pt]{pseudo-scalar}
               &~ $\eta^{PV}$ &~~&$\left( f_\eta^{B}/m_\eta \right)^2$        &~~  &$\left( M^{\ast2}+\varepsilon^\ast_k\varepsilon^\ast_p+\bm{k}\cdot\bm{p}\right)\bm{q}^2$\\
 \hline\hline
 \end{tabular}
 \end{table*}

In the following calculations, Bonn-B potential~\cite{MachleidtANP1989} will be adopted for $v_{pp}(\bm{k},\bm{p})$ and the effective interaction PK1~\cite{LongPK1PRC2004} of the relativistic mean-field theory is used in the $ph$ channel.

\section{Results and discussion}\label{Chapter3}

In the following discussion, the pure neutron matter will be mainly selected to clarify the physics.
To simulate the medium polarization effects and take into account the possible ambiguity of pairing force~\cite{CaoPRC2006}, the effective pairing interaction factor $\lambda$ varying from $0.8$ to $1.4$ is taken in the calculations. When the pairing gaps at the Fermi surface, $\Delta_{\rm{Fn}}\equiv\Delta(k_{\rm{Fn}})$,
are plotted as a function of the Fermi momentum $k_{\rm{Fn}}$ for different effective factor $\lambda$, a maximum $\Delta(k_{\rm{Fn}})$
always appears around $k_{\rm{Fn}}= 0.8~{\rm fm^{-1}}$ but changes from about $1.4~\rm{MeV}$ for $\lambda=0.8$ to about $6.5~\rm{MeV}$ for $\lambda=1.4$, and is $3.1~\rm{MeV}$ for $\lambda=1.0$, i.e., the relativistic Bonn-B potential.

For di-neutron correlations in the low density region, the scattering length $a$ in the $^{1}S_{0}$ channel, which is defined in terms of the $\rm{T}$-matrix for the scattering in the free space, is an important physical quantity. According to the definition, the negative value of the scattering length $a$ indicates an unbound state of neutron Cooper pairs while the positive value represents a bound state. It has been proposed to define the boundaries of the BCS-BEC crossover using a regularized gap equation approach~\cite{LeggettJPhysC1980, SadeMeloPRL1993,EngelbrechtPRB1997, MatsuoPRC2006}, which is related with the scattering length $a$. This approach is generically applied to dilute systems, where the interaction matrix elements $v_{pp}({\bm {k}},{\bm{ p}})$  can be approximately treated as constant $v_{0}$. Using the zero-energy $\rm{T}$ matrix, the relation between the scattering length $a$ in the $^{1}S_{0}$ channel and the constant interaction $v_{0}$ can be obtained as

\begin{equation}
\frac{M}{4\pi\hbar^{2}a}=\frac{1}{v_{0}}+\frac{1}{(2\pi)^{3}}\int d{\bm{ k}}\frac{1}{2e(k)},
\label{Eq:sca.length1}
\end{equation}
where $e(k)$ is the neutron kinetic energy,
\begin{equation}
e(k)=\sqrt{k^{2}+M^{*2}}-M^{*}.
\label{Eq:neu.kine.energy}
\end{equation}
The pairing gap equation (\ref{Eq:GapEq}) becomes
\begin{equation}
1 = -\frac{1}{(2\pi)^{3}}\int v_{0}\cdot\frac{1}{2E_{k}} d\bm{k}.
\label{Eq:red.gap.equ.}
\end{equation}
From Eq.~(\ref{Eq:sca.length1}) and Eq.~(\ref{Eq:red.gap.equ.}), the regularized gap equation is written as
\begin{equation}
 \frac{M}{4\pi\hbar^{2}a}=-\frac{1}{2(2\pi)^{3}}\int d\bm{k}\left[\frac{1}{E_k}-\frac{1}{e(k)}\right].
 \label{Eq:sca.length}
\end{equation}
Here instead of fixing the scattering length $a$ from its physical value, it is treated as a variable calculated by Eq.~(\ref{Eq:sca.length}) for different densities similarly as in Ref.~\cite{MatsuoPRC2006}. The quasi-particle energy $E_k$ and the neutron kinetic energy $e(k)$ in Eq.~(\ref{Eq:sca.length}) are calculated by the momentum dependent pairing gap $\Delta(k)$ and the effective mass $M^*$, which are obtained from the self-consistent RHB theory with Bonn potential for the pairing force. This is different from the regularized contact interaction model~\cite{MatsuoPRC2006}, where constant pairing gap is used. However, the difference should be small in the low-density limit and comparable results are expected.

\begin{figure}[t]
\includegraphics[width=0.5\textwidth]{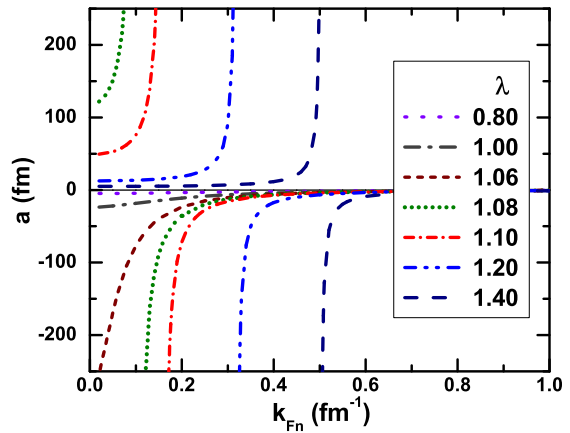}
\caption{(Color online) The neutron-neutron scattering length $a$ in the $^{1}S_{0}$ channel as a function of the neutron Fermi momentum $k_{\rm{Fn}}$ for different effective pairing interaction factors $\lambda$
in the pure neutron matter.}
\label{Fig:fig1}
\end{figure}

In Fig.~\ref{Fig:fig1}, the neutron-neutron scattering length $a$ in the $^{1}S_{0}$ channel is shown as a function of the neutron Fermi momentum $k_{\rm{Fn}}$ for different effective pairing interaction factors $\lambda$ in the pure neutron matter. When $\lambda \leqslant 1.06$, only negative branch of the scattering length appears and the scattering length approaches zero at high densities, suggesting no bound state for neutron Cooper pairs. However, when $\lambda \geq 1.08$, the positive branch of the scattering length becomes available at the dilute density, which implies the occurrence of a possible di-neutron bound state. For a given effective pairing interaction factor of $\lambda \geq 1.08$, with the increasing density, the scattering length starts from positive value and then diverges to positive infinity, and after crossing the Feshbach resonance, it reemerges from negative infinity. With increasing $\lambda$, the Feshbach resonance is shifted to higher density.

From the regularized gap equation (\ref{Eq:sca.length}), it was shown~\cite{MatsuoPRC2006, EngelbrechtPRB1997, MariniEPJB1998} that the properties of pairing correlations can be uniquely controlled by a dimensionless parameter $1/(k_{\rm{Fn}}a)$ which can give the evolution from BCS to BEC. The dimensionless parameter $1/(k_{\rm Fn}a)\ll-1$ corresponds to the weak coupling BCS regime, while $1/(k_{\rm Fn}a)\gg1$ is related to the strong correlated BEC regime. From weak coupling to strong attraction, the parameter $1/(k_{\rm Fn}a)$ evolves smoothly from negative to positive. The boundaries characterizing the BCS-BEC crossover can be approximately determined by $1/(k_{\rm Fn}a)=\pm1$~\cite{SadeMeloPRL1993, Randeria1995, EngelbrechtPRB1997}. The unitarity limit is defined as $1/(k_{\rm{Fn}}a)=0$, which is the midpoint of the BCS-BEC crossover.

\begin{figure}[t]
\includegraphics[width=0.5\textwidth]{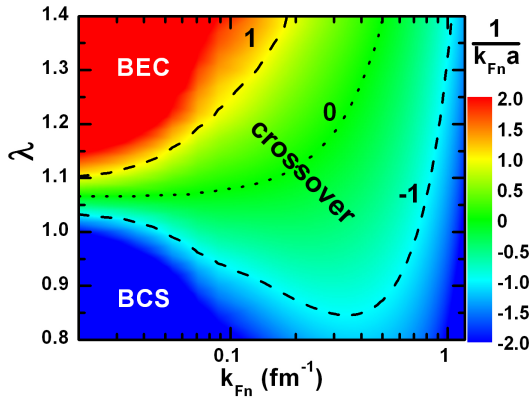}
\caption{(Color online) A contour plot for the dimensionless parameter $1/(k_{\rm{Fn}}a)$ as a function of the neutron Fermi momentum $k_{\rm{Fn}}$ and the effective factor $\lambda$ of pairing interaction in the pure neutron matter. The boundaries characterizing the BCS-BEC crossover ($1/(k_{\rm Fn}a)=\pm1$) are denoted by the two dashed lines, while the unitarity limit between BCS and BEC regime ($1/(k_{\rm{Fn}}a)=0$) is shown by the dotted line.}
\label{Fig:fig2}
\end{figure}

In Fig.~\ref{Fig:fig2}, a contour plot for the dimensionless parameter $1/(k_{\rm{Fn}}a)$ of the neutron Cooper
pairs is shown, as a function of the neutron Fermi momentum ~$k_{\rm{Fn}}$~and the effective pairing interaction factor~$\lambda$ in the pure neutron matter. The boundaries of the BCS-BEC crossover ($1/(k_{\rm Fn}a)=\pm1$) are denoted by the dashed lines, while the unitarity limit ($1/(k_{\rm{Fn}}a)=0$) is shown by the dotted line. Distinct features of di-neutron correlations are revealed for different pairing strengths.

At low density region with $k_{\rm{Fn}}\lesssim 0.2~{\rm fm^{-1}}$, the neutron Cooper pairs evolve continuously with the pairing strength from weak coupling BCS state to strong correlated BEC state.
The effective factor $\lambda$ corresponding to $1/(k_{\rm{Fn}}a)=1$ which characterizes the BEC boundary
starts around $\lambda=1.10$ in the low-density limit and increases with the density.
The effective factor $\lambda$ corresponding to $1/(k_{\rm{Fn}}a)=0$ which characterizes
the unitarity limit starts around 1.07 in the low-density limit and increases with the density monotonically.
The effective factor $\lambda$ corresponding to $1/(k_{\rm{Fn}}a)=-1$ which characterizes the BCS boundary
starts around $\lambda = 1.03$ in the low-density limit, decreases
with the density to a minimum  $\lambda \sim 0.85$ at $k_{\rm{Fn}}\thicksim0.35~\rm{fm}^{-1}$,
then increases sharply.

Besides the dimensionless parameter $1/(k_{\rm{Fn}}a)$, the density correlation function $D(q)$~\cite{BogdanPRL2005, IsayevPRC2008}, which describes the difference  between the mean field and the pairing field, is also a useful quantity to study the dependence of di-neutron correlations on the strength of pairing force. This measure
can give the transition points from BCS state to BEC state but not the BCS-BEC crossover region.
At zero-momentum transfer, i.e., $q=0$, the density correlation function $D(q=0)$ is reduced as
\begin{equation}
D(0)=\frac{1}{\pi^{2}\rho_{n}}\int_{0}^{\infty}\left(\kappa_{k}^{2}-\rho_{k}^{2}\right)k^{2}dk,
\label{Eq:D(0)}
\end{equation}
where $\kappa_k$ and $\rho_k$ are from Eq.(\ref{Eq:rho.kappa}).
The sign change of $D(0)$ has been considered as a criterion of the BCS-BEC crossover~\cite{BogdanPRL2005, IsayevPRC2008},
i.e., $D(0)<0$ means a BCS-type pairing and $D(0)>0$ represents a di-neutron BEC state.

\begin{figure}[t]
\includegraphics[width=0.5\textwidth]{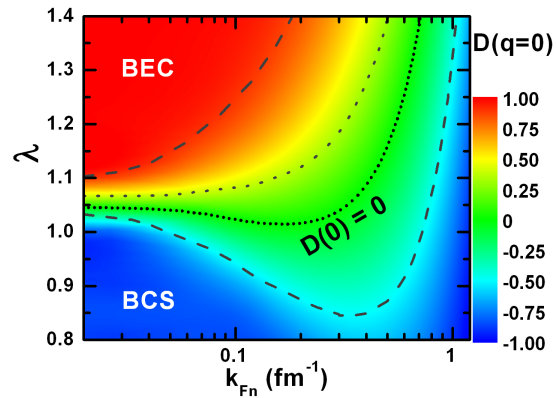}
\caption{(Color online) A contour plot for the density correlation function at zero-momentum transfer $D(0)$ as a function of the neutron Fermi momentum $k_{\rm{Fn}}$ and the effective factor $\lambda$ of pairing interaction in the pure neutron matter. The boundary where $D(0)=0$ is denoted by the short dotted line. For comparison, the two dashed lines characterizing the BCS-BEC crossover and the dotted line characterizing the unitarity limit between BCS and BEC regime in Fig.~\ref{Fig:fig2} are plotted as well.  }
\label{Fig:fig3}
\end{figure}

In Fig.~\ref{Fig:fig3}, the density correlation function at zero-momentum transfer $D(0)$ of the neutron Cooper pairs is shown as a function of the neutron Fermi momentum $k_{\rm{Fn}}$ and the effective pairing interaction factor $\lambda$ for the pure neutron matter. The critical line $D(0)=0$ is denoted by the short dotted line in comparison with the reference lines of $1/(k_{\rm{Fn}}a)=\pm1,0$ obtained from Fig.~\ref{Fig:fig2}.

In Fig.~\ref{Fig:fig3}, it is revealed that the density correlation function $D(0)$ has a similar pattern as the dimensionless parameter $1/(k_{\rm{Fn}}a)$. The effective factor $\lambda$ corresponding to $D(0)=0$ which characterizes the transition from BCS state to BEC state starts around $\lambda = 1.05$ in the low-density limit, decreases with the density to a minimum $\lambda \sim 1.02$ at $k_{\rm{Fn}}\thicksim0.20~\rm{fm}^{-1}$, then increases rapidly with the density. For $\lambda=1.0$, the results show that no di-neutron BEC state could occur, which agrees with previous results~\cite{SunPLB2010}. By taking the density correlation function $D(0)=0$ as a measure, the effective factor $\lambda$ characterizing the transition from BCS state to BEC state is smaller than the case by taking the unitarity limit $1/(k_{\rm{Fn}}a)=0$ as a measure.

In the case of the RMF theory, it has been proved that the neutron pair wave function in momentum space $\Psi_{\rm pair}(k)$, i.e., the anomalous density $\kappa_{k}$ in Eq.~(\ref{Eq:rho.kappa}), satisfies a Schr\"{o}dinger-like equation which is expressed as~\cite{SunPLB2010}
\begin{eqnarray}
&&2e(k)\Psi_{\rm{pair}}(k)+\frac{1-2\rho_{k}}{4\pi^{2}}\int_{0}^{\infty}\lambda\cdot v_{pp}(k,p)p^{2}dp\Psi_{\rm{pair}}(p)\nonumber\\
&&=2\nu_{\rm{n}}\Psi_{\rm{pair}}(k),
\label{Eq:nu.sch}
\end{eqnarray}
with the corresponding energy eigenvalue $2\nu_{\rm{n}}$, where $\nu_{\rm{n}}$ is the effective neutron chemical potential obtained by deducting the momentum independent part from the chemical potential $\mu_{\rm{n}}$,
\begin{equation}
\nu_{\rm{n}}=\mu_{\rm{n}}-\Sigma_{0}-M^{*}.
\label{Eq:nu}
\end{equation}
In the limit of zero density, the effective chemical potential $\nu_{\rm n}$ behaves as half binding energy of the Cooper pair~\cite{NozieresJLTP1985}. For the evolution from the weak coupling BCS regime to the strongly correlated BEC regime, the effective chemical potential $\nu_{\rm{n}}$ is supposed to change from positive to negative.

\begin{figure}[t]
\includegraphics[width=0.5\textwidth]{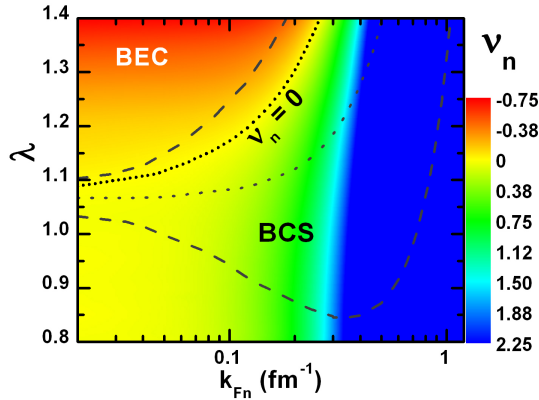}
\caption{(Color online) A contour plot for the effective neutron chemical potential~$\nu_n$~as a function of the neutron Fermi momentum $k_{\rm{Fn}}$ and the effective factor $\lambda$ of pairing interaction in the pure neutron matter. The boundary of $\nu_{n}=0$ is denoted by the short dotted line. For comparison, the two dashed lines characterizing the BCS-BEC crossover and the dotted line characterizing the unitarity limit between BCS and BEC regime in Fig.\ref{Fig:fig2} are plotted as well.}
\label{Fig:fig4}
\end{figure}

In Fig.~\ref{Fig:fig4}, a contour plot for the effective chemical potential $\nu_{\rm{n}}$ of the neutron Cooper pair is shown as a function of $k_{\rm{Fn}}$ and $\lambda$ in the pure neutron matter. The short dotted line denotes $\nu_{\rm{n}}=0$ in comparison with the dimensionless parameter $1/(k_{\rm{Fn}}a)=\pm1, 0$ extracted from Fig.~\ref{Fig:fig2}.

The effective factor $\lambda$ corresponding to
$\nu_{\rm{n}}=0$ which characterizes the transition from
BCS state to BEC state starts around $\lambda=1.09$
in the low-density limit, then increases monotonically with the
density. For $\lambda=1.0$, there is no evidence for the
BEC state of neutron pairs at any density, similarly as the
case by taking density correlation function $D(0)=0$ as a
measure. By taking the effective chemical potential
$\nu_{\rm{n}}=0$ as a measure, the effective factor
$\lambda$ characterizing the transition from BCS state to
BEC state is larger than the case by taking the unitarity
limit $1/(k_{\rm{Fn}}a)=0$ as a measure.

Combining the conclusions mentioned above, it can be summarized that a di-neutron BEC state will appear in the low-density limit for $\lambda \geq 1.10$ and there is only a weak coupling BCS state for $\lambda\lesssim 0.85$.
For $\lambda = 1.10$, the maximum pairing gap at the Fermi surface $\Delta_{\rm{Fn}}$ is $4.12~\rm{MeV}$ around the density $k_{\rm{Fn}}=0.8~{\rm fm^{-1}}$, and for $\lambda = 0.85$, it is $1.78~\rm{MeV}$.

For symmetric nuclear matter, recent studies have claimed that a di-neutron BEC state can be formed at $k_{\rm{Fn}}\sim 0.2 ~\rm{fm}^{-1}$ after considering the medium polarization effects ~\cite{MargueronSagawaPRC2007, IsayevPRC2008}. In order to examine the above conclusion, similar investigation as the pure neutron matter for the symmetric nuclear matter has been done.

For symmetric nuclear matter, by taking the dimensionless parameter $1/(k_{\rm{Fn}}a)$ as a measure, a di-neutron BEC state will occur in the low-density limit for $\lambda \geq 1.10$ and neutron Cooper pairs are totally in the BCS state for $\lambda \lesssim 0.85$. By taking the density correlation function with zero-momentum transfer $D(0)$ as a measure, the di-neutron BEC state will appear in the low-density limit for $\lambda \gtrsim 1.05$. The minimum effective factor $\lambda$ corresponding to $D(0)=0$ is around $1.01$ with the neutron Fermi momentum $k_{\rm Fn}\sim 0.2~{\rm fm}^{-1}$. By taking the effective chemical potential $\nu_{\rm n}=0$ as a measure, the
effective factor $\lambda$ characterizing the transition from BCS state to BEC state is larger than
the case by taking the density correlation function $D(0)=0$ as a measure and the di-neutron BEC state may occur in the low-density limit for $\lambda \gtrsim 1.08$. Combining the conclusions mentioned above, for symmetric nuclear matter, a di-neutron BEC state will appear in the low-density limit for $\lambda \geq 1.10$ and there is only a weak coupling BCS state for $\lambda\lesssim 0.85$, which are similar as the pure neutron matter.

\begin{table*}
\tabcolsep=5pt
\caption{Reference values of $P(d_{\rm n})$, $\xi_{\rm rms}/d_{\rm n}$, $\Delta_{\rm Fn}/e_{\rm Fn}$ and $\nu_{\rm n}/e_{\rm Fn}$ characterizing the BCS-BEC crossover in the pure neutron matter (symmetric nuclear matter) respectively for the dimensionless parameter $1/(k_{\rm Fn}a) = 0, \pm 1$, the zero-momentum transfer density correlation function $D(0)=0$, and the effective chemical potential $\nu_{\rm n}=0$ with effective pairing force factor $\lambda=1.0, 1.1, 1.2$. For comparison, the results obtained in the regularized contact interaction model~\cite{MatsuoPRC2006, EngelbrechtPRB1997} are also shown.}
\label{Tab:CQNM}
\begin{tabular}{cccccccccc}
\hline\hline
       &&&\multicolumn{3}{c} {$1/(k_{\rm{Fn}}a)$}&&{D(0)}&&{$\nu_{n}$}\\
        \cline{4-6}\cline{8-8}\cline{10-10}
          &\raisebox{2.3ex}[0pt]{~$\lambda$~}&& -1 & 0 & +1 & & 0 &  &  0\\
\hline
                &1.0   && 0.82 (0.81) & ！         &  ！         &&  ！        && ！\\
                &1.1   && 0.82 (0.81) & 0.99 (1.00) &  ！         && 0.95 (0.95) && ！\\
\raisebox{2.3ex}[0pt]{$P(d_{\rm{n}})$}
                &1.2   && 0.82 (0.81)& 0.99 (0.99) & 1.00 (1.00)  && 0.96 (0.96) && 1.00(1.00)\\
                &\cite{MatsuoPRC2006}&& 0.81 & 0.99 & 1.00  &&  ！  && ！\\
\hline
                &1.0&& 0.98 (1.00)   & ！ & ！    && ！ && ！\\
                &1.1&& 0.98 (1.03)    & 0.35 (0.35) & ！    && 0.52 (0.51) && ！\\
\raisebox{2.3ex}[0pt]{$\xi_{\rm{rms}}/d_{\rm{n}}$}
                &1.2&& 0.98 (1.04) & 0.38 (0.38) & 0.19 (0.21)  && 0.52 (0.51) && 0.26 (0.25)\\
                &\cite{MatsuoPRC2006}&&1.10 & 0.36 & 0.19  &&  ！  && ！\\
\hline
                &1.0&& 0.25 (0.25)   & ！  & ！    && ！ && ！\\
                &1.1&& 0.26 (0.27)   & 0.74 (0.78)  & ！    && 0.56 (0.55) && ！\\
\raisebox{2.3ex}[0pt]{$\Delta_{\rm{Fn}}/e_{\rm{Fn}}$}
                &1.2&& 0.28 (0.26) & 0.83 (0.82) & 1.52 (1.50)  && 0.59 (0.59) && 1.20 (1.21)\\
                &\cite{MatsuoPRC2006}&&0.21 & 0.69 & 1.33  &&  ！  && ！\\
\hline
                &1.0&& 0.99 (0.98) & ！ &！  && ！ && ！\\
                &1.1&& 0.98 (0.98) & 0.61 (0.61) &！  && 0.85 (0.84) && ！\\
\raisebox{2.3ex}[0pt]{$\nu_{\rm{n}}/e_{\rm{Fn}}$}
                &1.2 && 0.99 (0.99) & 0.64 (0.63) &-0.84 (-0.87)  && 0.85 (0.85) && -0.01 (-0.01)\\
                &\cite{EngelbrechtPRB1997}&& 0.97 & 0.60 &-0.77  &&  ！  && ！\\

\hline\hline
\end{tabular}
\end{table*}

In Refs.~\cite{MatsuoPRC2006, MargueronSagawaPRC2007}, the BCS-BEC crossover is investigated by several characteristic quantities including the probability for the neutron pair partners $P(d_{\rm{n}})$ as well as the ratios $\xi_{\rm{rms}}/d_{\rm{n}}$, $\Delta_{\rm{Fn}}/e_{\rm{Fn}}$, and $\nu_{\rm{n}}/e_{\rm{Fn}}$, where $\Delta_{\rm{Fn}}$ is the neutron pairing gap at the Fermi surface, $e_{\rm{Fn}}$ the neutron Fermi kinetic energy in Eq.~(\ref{Eq:neu.kine.energy}) with $e_{\rm Fn} = e(k=k_{\rm Fn})$, $\nu_{\rm{n}}$ the effective chemical potential, $P(d_{\rm{n}})=\int\limits_{0}^{d_{\rm{n}}}|\Psi_{\rm{pair}}(r)|^{2}r^{2}dr$
with the average inter-neutron distance $d_{\rm{n}}\equiv \rho_{\rm{n}}^{-1/3}$ and $\Psi_{\rm{pair}}(r)$ the neutron Cooper pair wave function in coordinate space, and the mean square radius of the neutron Cooper pairs $\xi_{\rm{rms}}^{2}= {\int|\Psi_{\rm{pair}}(r)|^{2}r^{4}dr} / {\int|\Psi_{\rm{pair}}(r)|^{2}r^{2}dr}$. As these characteristic quantities are monotonic functions of the dimensionless parameter $1/(k_{\rm{Fn}}a)$ in the regularized gap equation approach~\cite{MatsuoPRC2006, EngelbrechtPRB1997, MariniEPJB1998}, they can be used to describe the boundaries of BCS-BEC crossover.

In Table \ref{Tab:CQNM} are respectively listed the values of $P(d_{\rm n})$, $\xi_{\rm rms}/d_{\rm n}$, $\Delta_{\rm Fn}/e_{\rm Fn}$ and $\nu_{\rm n}/e_{\rm Fn}$ characterizing the BCS-BEC crossover in the pure neutron matter for the dimensionless parameter $1/(k_{\rm Fn}a)= 0,\pm 1$, the zero-momentum transfer density correlation function $D(0)=0$, and the effective chemical potential $\nu_{\rm n}=0$ with effective pairing factors $\lambda=1.0, 1.1, 1.2$. The corresponding values for symmetric nuclear matter are given in the parenthesises. For comparison, the values in the regularized contact interaction model~\cite{MatsuoPRC2006, EngelbrechtPRB1997} are also given.

In Table \ref{Tab:CQNM}, it is shown that the values of $P(d_{\rm n})$, $\xi_{\rm rms}/d_{\rm n}$, and $\nu_{\rm n}/e_{\rm Fn}$ are almost independent of the pairing interaction strength and are similar for the pure neutron matter and the symmetric nuclear matter. Furthermore, the values of $P(d_{\rm n})$, $\xi_{\rm rms}/d_{\rm n}$, and $\nu_{\rm n}/e_{\rm Fn}$ are consistent with the results obtained in the regularized contact interaction model ~\cite{MatsuoPRC2006, EngelbrechtPRB1997}. In Table \ref{Tab:CQNM}, the values of $\Delta_{\rm Fn}/e_{\rm Fn}$ slightly increase with the pairing interaction strength, which might be ascribed to the fact that the pairing gap $\Delta_{\rm Fn}$ increases with the pairing force strength faster than the neutron Fermi kinetic energy $e_{\rm Fn}$ does. Furthermore, difference between the present $\Delta_{\rm Fn}/e_{\rm Fn}$ and those in the regularized contact interaction model~\cite{MatsuoPRC2006} exists, which might be ascribed to the pairing force, i.e., the Bonn potential here and the constant interaction in the regularized contact interaction model.
As the reference values of the characteristic quantities $P(d_{\rm n})$, $\xi_{\rm rms}/d_{\rm n}$, $\Delta_{\rm Fn}/e_{\rm Fn}$ and $\nu_{\rm n}/e_{\rm Fn}$ characterizing the boundaries of BCS-BEC crossover are obtained in the relativistic framework by a self-consistent way and are consistent with those got by the regularized contact interaction model of non-relativistic framework in Refs.~\cite{MatsuoPRC2006, EngelbrechtPRB1997}, they provide a valuable guide in characterizing BCS-BEC crossover boundaries in future investigations.

\section{Conclusion}\label{Chapter4}
In conclusion, the influence of the pairing interaction strength on the di-neutron correlations in the $^1S_0$ channel and the BCS-BEC crossover phenomenon in nuclear matter has been investigated based on the RHB theory. The effective interaction PK1 is adopted in the $ph$ channel and the Bonn-B potential is used in the $pp$ channel. The influence of medium polarization effects on the pairing properties and the possible ambiguity of pairing force are simulated by an effective factor $\lambda$ appending on the Bonn-B potential.

From the dimensionless parameter $1/(k_{\rm Fn}a)$, the zero-momentum transfer density correlation function $D(0)$ and the effective chemical potential $\nu_{\rm n}$, a di-neutron BEC state will occur at dilute density if $\lambda\geq1.10$, and there is only a BCS state if $\lambda\lesssim0.85$. Moreover, the reference values of several characterized quantities $P(d_{\rm n})$, $\xi_{\rm rms}/d_{\rm n}$, $\Delta_{\rm Fn}/e_{\rm Fn}$ and $\nu_{\rm n}/e_{\rm Fn}$ characterizing the boundaries of BCS-BEC crossover are obtained in the self-consistent relativistic framework and are consistent with the non-relativistic results~\cite{MatsuoPRC2006, EngelbrechtPRB1997}, which may provide a valuable guide in characterizing BCS-BEC crossover boundaries in future investigations.

\begin{acknowledgments}
We would like to thank M. Matsuo and H. Toki for valuable comments and discussions. This work was partly supported by the Major State 973 Program of China (Grant No. 2007CB815000), the Natural Science Foundation of China (Grant Nos. 10975008, 11175002), the Fundamental Research Funds for the Central Universities (Grant Nos. lzujbky-2012-k07 and No. lzujbky-2012-07), and the Research Fund for the Doctoral Program of Higher Education under Grant No. 20110001110087.
\end{acknowledgments}

\end{document}